\documentstyle[preprint,aps]{revtex}
\begin{document} 
\draft
\title{Topics Concerning the Quadrupole-Quadrupole Interaction}
 
\author{M.S. Fayache$^1$, Y.Y. Sharon$^2$ and
L. Zamick$^3$\cite{byline}\\  
(1) D\'{e}partement de Physique, Facult\'{e} des Sciences de Tunis\\
Tunis 1060, Tunisia\\
\noindent (2) Department of Physics and Astronomy, Rutgers University,
Piscataway, New Jersey 08855\\
\noindent (3) TRIUMF, 4004 Wesbrook Mall\\
Vancouver, B.C., CANADA V6T 2A3\\}
\date{\today}
\maketitle

\begin{abstract}
We address some properties of the quadrupole-quadrupole ($Q \cdot Q$)
interaction in nuclear studies. We first consider how to restore
$SU(3)$ symmetry even though we use only coordinate and not momentum
terms. Using the Hamiltonian $H=\sum_i (\frac{p^2}{2m} +
\frac{m}{2}\omega^2 r_i^2) 
-\chi \sum_{i < j}Q(i) \cdot Q(j) -\frac{\chi}{2} \sum_i Q(i) \cdot Q(i)$ with
$Q_{\mu}=r^2 Y_{2,\mu}$, we find that only 2/3 of the single-particle
splitting ($\epsilon_{0d}-\epsilon_{1s}$) comes from the diagonal term
of $Q \cdot Q$ -the remaining 1/3 comes from the interaction of the
valence nucleus 
with the core. On another topic, a previously derived relation, using
$Q \cdot Q$, between isovector orbital $B(M1)$ (scissors mode) and the
{\em difference} ($B(E2, isoscalar)-B(E2, isovector)$) is
discussed. It is shown that one needs the isovector $B(E2)$ in order
that one get the correct limit as one goes to nuclei sufficiently far
from stability so that one subshell (neutron or proton) is closed. 
\end{abstract}

In this work we address issues pertaining to shell model calculations
with the schematic quadrupole-quadrupole interaction. Even today, this
interaction is of value in casting light upon the relationship between
shell model and collective model behaviour. There are still new things
to be learnt about this interaction in nuclei, and we will discuss two
examples here. 

\section{The Single-Particle Splitting ($\epsilon_{0d}-\epsilon_{1s}$)
Needed to Get the $SU(3)$ Result}

We wish to obtain Elliott's $SU(3)$ results \cite{Elliott} in a shell
model calculation in which only the coordinate $Q \cdot Q$ interaction
is used. We do not wish to use the momentum-dependent terms. The
latter were introduced by Elliott so that, in combination with the
coordinate terms, there would be no $\Delta N=2$ admixtures i.e. no
admixture from configurations involving 2 $\hbar \omega$
excitations. However, we {\em want } to see the effects of such
admixtures in our shell model studies. One classic problem in which
$\Delta N=2$ admixtures are important id the $E2$ effective charge,
but there are many other problems of interest along these lines.

The Hamiltonian we consider is therefore

\[H=\sum_i (\frac{p^2}{2m} + \frac{1}{2}m \omega^2 r_i^2)
-\chi \sum_{i < j}Q(i) \cdot Q(j) -\frac{\chi}{2} \sum_i Q(i) \cdot
Q(i)\] 

\noindent where $Q(i)^k \cdot Q(j)^k=(-1)^k \sqrt{2k+1} r(i)^kr(j)^k
[Y(i)^k Y(j)^k]^0$ with $k=2$. Like Elliott, we have not only the
two-body $Q \cdot Q$ term, but also the $i=j$ single-particle term. 

It is convenient to introduce the following quantity: $\bar{\chi}=5
b^4 \chi/32 \pi$ where $b$ is the oscillator length parameter, such
that $b^2=\hbar/m \omega=41.46/\hbar \omega$. 

To evaluate the single-particle term we use the addition theorem:

\[ P_k(cos \theta_{12})=\frac{4 \pi}{2k+1}\sum_{\mu}Y_{k,\mu}(1) Y_{k,
-\mu}(2)\] 

\noindent thus

\[\sqrt{5}[Y^2(i)Y^2(i)]^0=\frac{5}{4 \pi}P_2(1)=\frac{5}{4 \pi}\]

\noindent The single-particle potential is then 

\[ U(r)=-\frac{\chi}{2} Q(i) \cdot Q(i)= =4 \bar{\chi}\left(
\frac{r}{b} \right)^4\]

The expectation value of this single-particle term for various
single-particle states is given in Table I. What single-particle
splitting $\epsilon_{0d}-\epsilon_{1s}$ is needed to get Elliott's
$SU(3)$ results? The best way to answer this is to give the formula
for the $SU(3)$ energy in the $1s-0d$ shell (in which the momentum
terms are included):

\[E(\lambda \mu)=\bar{\chi}\left[-4(\lambda^2+\mu^2+\lambda\mu+ 3
(\lambda+\mu))+3L(L+1) \right] \]

\noindent For a rotational band, the $L=2 - L=0$ splitting is given by
the last term and is equal to $18 \bar{\chi}$. This must also be the 
$\epsilon_{0d}-\epsilon_{1s}$ because it is also an $L=2 - L=0$
splitting. But, as seen from Table I, the splitting due to the
diagonal $Q \cdot Q$ interaction is
$(-63-(-75))\bar{\chi}=12\bar{\chi}$. Where does the remaining
$6\bar{\chi}$ come from?

The answer is that the missing part comes from the interaction of the
particle with the core. For $Q \cdot Q$, the only
contribution is the exchange term of the $0d$ particle with the $0s$
core. 

Thus, to get the Elliott $SU(3)$ results in the $1s-0d$ shell, we must
not only include his diagonal term but also include the particle-core
interaction {\em i.e.} take the shell model as an $A$ particle problem
rather than an ($A-16)$ particle problem. 

The same thing happens in the $0f-1p$ shell. The single-particle
splitting required to get the $SU(3)$ result is 
$\epsilon_{0f}-\epsilon_{1p}=3(3\times4-1\times2)\bar{\chi}=30\bar{\chi}$.
As seen from Table I, we only get 2/3 of this (20$\bar{\chi}$) from
the diagonal $Q \cdot Q$ term. The remaining 10$\bar{\chi}$ comes from
the interaction of the valence nucleons with the core (actually only
the $0p$ shell in the core will contribute).

\section{Clarification of a Relation Between the Isovector Orbital
Magnetic Dipole Transition Rate (i.e. Scissors Mode Exciation Rate)
and the Electric Quadrupole Transition.}

As a second example, we will attempt to clarify a relationship between
orbital magnetic dipole transition rates (i.e. scissors mode
excitation rates) and electric quadrupole transitions. 
Using the interaction $-\chi Q \cdot Q$, Zheng and Zamick \cite{zz}
obtained a sum rule relating these two quantities. The isovector
orbital magnetic dipole operator is ($\vec{L_{\pi}}-\vec{L_{\nu}}$)/2
(the isoscalar one is half the total orbital angular momentum
$\vec{L}/2=(\vec{L_{\pi}}+\vec{L_{\nu}})/2$). In detail, the sum rule
reads

\begin{eqnarray*}
\sum_n (E_n-E_0)B(M1)_o=\frac{9\chi}{16\pi}~\sum_i \left\{[B(E2,0_1 
\rightarrow 2_i)_{IS} \right. \nonumber\\
\left. - B(E2,0_1 \rightarrow 2_i)_{IV}] \right\}\\
\end{eqnarray*}
where $B(M1)_o$ is the value for the {\em isovector} orbital $M1$
operator ($g_{l\pi}=0.5$ $g_{l\nu}=-0.5$ $g_{s\pi}=0$ $g_{s\nu}=0$)
and the operator for the $E2$ transitions is $\sum_{protons}e_pr^2Y_2$ 
$+$ $\sum_{neutrons}e_nr^2Y_2$ with $e_p=1$, $e_n=1$ for the isoscalar
transition ($IS$), and $e_p=1$, $e_n=-1$ for the isovector transition
($IV$). The above result holds also if we add a pairing interaction
between like particles $i.e.$ between two neutrons and between two
protons. 

The above work was motivated by the realization from many sources that
there should be a relation between the scissors mode excitation rate
and nuclear collectivity. Indeed, the initial picture by Palumbo and
LoIudice \cite{ip} was of an excitation in a deformed nucleus in which
the symmetry axis of the neutrons vibrated against that of the
protons. In a 1990 contribution by the Darmstadt group \cite{rich}, it
was noted that 
the $Sm$ isotopes, which undergo large changes in deformation, the
$B(M1)_{scissors}$ was proportional to $B(E2,0_1 \rightarrow
2_1)$. The $B(E2)$ in turn is proportional to the square of the
nuclear deformation $\delta^2$. 

The above energy-weighted sum rule of Zheng and Zamick was an attempt
to obtain such a relationship microscopically using fermions rather
than interacting bosons. To a large extent they succeeded, but there
are some differences. Rather than the proportionality factor $B(E2,0_1 
\rightarrow 2_1)$, there is the difference of the isoscalar and
isovector $B(E2)$. Now one generally expects the isoscalar $E2$ state
to be most collective and much larger than the isovector $B(E2)$. If
the latter is negligible, then indeed one basically has the same
relation between scissors mode excitations and nuclear collectivity,
as empirically observed in the $Sm$ isotopes. 

However, derivation of the above energy-weighted sum rule is quite
general, and should therefore hold (in the mathematical sense) in all
regions -not just where the deformation is strong. To best illustrate
the need for the isovector $B(E2)$, consider a nucleus with a closed
shell of neutrons or protons. In such a nucleus, and neglecting
ground-state correlations, the scissors mode excitation rate will
vanish -one needs both open shell neutrons and protons to get a finite
scissors mode excitation rate. On the other hand, the $B(E2,0_1 
\rightarrow 2_1)$ can be quite large. However, if we have say an open
shell of protons and a closed shell of neutrons, the $B(E2,0_1 
\rightarrow 2_1)$ can be quite substantial. Many vibrational nuclei
are of such an ilk, and they have large $B(E2)$'s from ground $e.g.$ ~
20 W.u. 

However, in the above circumstances, the neutrons will not contribute
to the $B(E2)$ even if we give them an effective charge. But if only
the protons contribute, it is clear that $B(E2,
isovector)=B(E2,isoscalar)$.

As an example, let us consider the even-even $Be$ isotopes $^6Be$,
$^8Be$, $^{10}Be$ and $^{12}Be$. In so doing, we go far away from the
valley of stability, but this is in tune with modern interests in
radioactive beams. 

Fayache, Sharma and Zamick \cite{qqt} have previously considered
$^8Be$ and $^{10}Be$. The point was made that these two nuclei had
about the same calculated $B(E2,0_1 \rightarrow 2_1)$, but the
isovector orbital $B(M1)$'s were significantly smaller in $^{10}Be$
than in $^8Be$. This was against the systematic that $B(M1)_{orbital}$
is proportional merely to $B(E2)$. In detail, the calculated $B(M1,0_1
\rightarrow 1_1)$ was $2/\pi \mu_N^2$ for $^8Be$, and in $^{10}Be$ it
was $9/32\pi \mu_N^2~(T=1 \rightarrow T=1)$ and $15/32\pi \mu_N^2~(T=1
\rightarrow T=2)$ . Thus the ratio of isovector orbital $B(M1)$'s 
$^{10}Be/^8Be=3/8$. 

We now extend the calculations to include $^6Be$ and $^{12}Be$. These
are singly closed nuclei. We see in Table II how everything hangs
together. We can explain the reduction in $B(M1)$ in $^{10}Be$
relative to $^8Be$ by the fact that the isovector $B(E2)$ in $^{10}Be$
is much larger than in $^8Be$. Note that the isoscalar $B(E2)$'s are
almost the same in these two nuclei. The summed $B(M1)$ in $^8Be$ is
$2/\pi$, but in $^{10}Be$ it is only $3/8$ of that. 

In $^6Be$ and $^{12}Be$, the $E2$ transition is from two protons with
$L=0~S=0$ to two protons with $L=2~S=0$. Note that, surprisingly, the
coefficients in front of the effective charge factors is larger for
singly-magic $^6Be$ than it is for the open shell nucleus $^8Be$. The
factors are respectively $12.5b^4/\pi$ and $8.75b^4/\pi$. However, the
charge factor for $^6Be$ ($^{12}Be$) is $e_p^2$, whereas for $^8Be$ it
is $(e_p+e_n)^2$. The latter gives a factor of four enhancement for
the isoscalar $B(E2)$ in $^8Be$.

Again we see from Table II that the isoscalar and isovector $B(E2)$'s
are necessarily the same and, when this is fed into the sum rule of
Zheng and Zamick \cite{zz}, one gets the consistent result that
$B(M1)_{orbital}$ is zero. 

This work was supported by a Department of Energy grant
DE-FG02-95ER 40940. One of us (L.Z.) thanks E. Vogt, B. Jennings, and
H. Fearing for their help and hospitality at TRIUMF.

\begin{table}
\caption{The Expectation Value of $U/\bar{\chi}=-4 (r/b)^4$ for several 
single-particle states}
\begin{tabular}{cc}
state & $\langle U/\bar{\chi}\rangle$\\
\tableline
$0s$ & -15\\
$0p$ & -35\\
$0d$ & -63\\
$1s$ & -75\\
$0f$ & -99\\
$1p$ & -119\\
\end{tabular}
\end{table}

\begin{table}
\caption{The Values of $B(M1)_{orbital}$ and $B(E2)_{isoscalar}$ and 
$B(E2)_{isosvector}$ for $Be$ isotopes.}
\begin{tabular}{ccccc}
Nucleus & $B(M1)_{orbital}$ &
$B(E2)_{isoscalar}~(e^2fm^4)$\tablenotemark[1] &
$B(E2)_{isosvector}~(e^2fm^4)$\tablenotemark[1]\\ 
\tableline
$^6Be$ & 0 & 19.92 \tablenotemark[2] & 19.92 \tablenotemark[2]\\ 
& & & & \\
$^8Be$ & $2/\pi$=0.637 & 82.66 \tablenotemark[3] & 7.371 \\
& & & & \\
$^{10}Be$ & $T=1~\rightarrow T=1$~~ $9/32\pi$=0.0895 & 70.78 
\tablenotemark[4] & 32.35 \tablenotemark[5]\\
          & $T=1~\rightarrow T=2$~~ $15/32\pi$=0.149 & 0 & 3.322\\
& & & & \\
$^{12}Be$ & 0 & 19.92 & 19.92\\
\end{tabular}
\tablenotetext[1] {The value $b=1.650~fm$ was used for all nuclei above.}
\tablenotetext[2] {The analytic expression in $^6Be$ is
$B(E2)=\frac{50}{4\pi}b^4e_p^2$.}
\tablenotetext[3] {The analytic expression in $^8Be$ is
$B(E2)=\frac{35}{4\pi}b^4(e_p+e_n)^2$.}
\tablenotetext[4] {$B(E2)_{isoscalar}=0$ to the $2^+_1$ state, and is
equal to 68.24 $e^2fm^4$ to the $2^+_2$ state.}
\tablenotetext[5] {$B(E2)_{isovector}=31.19 e^2fm^4$ to the $2^+_1$
state, and is equal to zero for the $2^+_2$ state.}
\end{table}
\end{document}